\pdfoutput=1
\documentclass[onecolumn]{openjournal}

\usepackage{graphicx}
\usepackage{amsmath}
\usepackage{amssymb}
\usepackage{booktabs}
\usepackage{url}
\usepackage{orcidlink}
\hypersetup{colorlinks=true,urlcolor=blue,citecolor=blue,linkcolor=blue}

\newcommand{\Palive}{P_{\rm alive}}
\newcommand{\ST}{S_{T}}
\newcommand{\Rearth}{R_\oplus}
\newcommand{\Mearth}{M_\oplus}
\newcommand{\Msun}{M_\odot}

\begin{document}

\title{Probabilistic, facility-specific prioritisation of temperate exoplanets for atmospheric characterisation with JWST and the ELT}

\author{Effy Viridis\,\orcidlink{0009-0004-5620-5352}}
\affiliation{Independent Researcher, El Salvador}

\begin{abstract}
Deciding which temperate exoplanets to observe first matters for JWST transit spectroscopy and for planned ELT reflected-light instruments, yet signal-to-noise metrics ignore habitability and similarity indices ignore observability. We present an open Monte Carlo framework (UHI v6.1) that builds probabilistic priority lists separately for each facility and composition class, never as a single ranking. For each of 6,112 planets in the NASA Exoplanet Archive, 10,000 draws carry catalogue uncertainties through climate, cumulative stellar-XUV atmospheric retention, per-draw rocky, Hycean or gas-rich classes, and a JWST (transiting) or ELT/PCS (non-transiting) detection model, giving each planet a survival probability, a conditional score and a top-N probability. At $5\sigma$, 255 rocky and 71 Hycean candidates remain viable. A Saltelli--Sobol analysis of 32 assumptions (N = 512) shows that the location and width of the XUV shoreline dominate the rocky lists, with total-order indices of 0.22 and 0.12 at $5\sigma$ (0.25 and 0.14 at $3\sigma$), and all ten leading indices have 95\% bootstrap half-widths below 0.05. For Hycean planets most outputs are degenerate and the analysis does not converge, so that branch is reported as indicative. Membership of these lists is limited mainly by whether a planet survives the habitability vetoes: at $5\sigma$, 3 of the nominal JWST and 4 of the nominal ELT rocky top-10 targets enter the top 10 in at least half of all draws, yet every one of them stays in the top 10 in more than half of the draws in which it survives. Code, per-planet posterior summaries, sensitivity tables and maps are public, so observers can rerun the lists under their own instrument assumptions and see which inputs would most change a target's priority.
\end{abstract}

\keywords{Exoplanets (498) --- Habitable planets (695) --- Exoplanet atmospheres (487) --- Transmission spectroscopy (2133) --- Direct imaging (387) --- Astrostatistics (1882) --- Computational methods (1965)}

\section{Introduction}
\label{sec:intro}

Telescope time for the atmospheres of small, temperate planets is scarce and will stay scarce. The STScI working group on strategic exoplanet initiatives projected that JWST could devote of order 30,000~hours to exoplanets over its lifetime and recommended a 500-hour Director's Discretionary programme aimed at rocky planets orbiting M dwarfs, with the explicit goal of locating the cosmic shoreline \citep{Redfield2024}. On the ground, the Planetary Camera and Spectrograph (PCS) concept for the ELT is designed to image nearby rocky planets in reflected light \citep{Kasper2021}, and its science case already depends on lists of known radial-velocity planets around the nearest stars. In both settings a proposer or a committee has to answer the same practical question: of the planets that could be observed, which ones deserve the hours first?

Existing tools address parts of that question. The transmission and emission spectroscopy metrics (TSM and ESM) of \citet{Kempton2018} estimate the signal a transiting planet would give JWST, and they have been used to sort candidates into bins of radius and equilibrium temperature for proposal planning \citep{Hord2024}. These metrics deliberately say nothing about whether a planet is temperate or whether it is likely to have kept an atmosphere. Habitability indices take the opposite view. The Earth Similarity Index \citep{SchulzeMakuch2011} compares bulk properties with those of the Earth, and the Statistical-likelihood Exo-Planetary Habitability Index, SEPHI \citep{RodriguezMozos2017}, combines likelihood functions for several physical criteria. Its recent extension, SEPHI~2.0 \citep{RodriguezMozos2025}, adds eccentricity, an updated mass--radius relation, estimated magnetic fields and a combined treatment of thermal (Jeans) and non-thermal (stellar-wind) atmospheric escape, takes the most restrictive criterion as the final index, propagates the input uncertainties with a Monte Carlo loop, and is offered through a public tool intended to help prioritise follow-up observations. Those indices were built to assess habitability potential, and they do that carefully; they do not model atmospheric loss driven by the cumulative XUV fluence, and they are not designed to say whether a given instrument can detect the planet's atmosphere. Facility-specific star lists, such as the NASA ExEP list of 164 stars for the Habitable Worlds Observatory \citep{Mamajek2024}, show how a stated observational purpose sharpens target selection. Closest to the present work, \citet{Ih2026} treat the choice of JWST rocky-planet targets as a combinatorial optimisation, a ``knapsack'' problem, and show how a well-chosen sample can test the cosmic-shoreline hypothesis statistically. Their framework optimises a survey design for a population-level question, whereas the lists presented here rank individual planets and expose the assumptions behind each rank, so the two approaches are complementary.

This paper describes version~6.1 of an open framework, distributed under the name UHI (Unified Habitability Index), that joins these threads into probabilistic priority lists for atmospheric follow-up. An earlier and much simpler version of the code, v5.2, was archived on Zenodo together with a short research note \citep[v5.2;][]{Viridis2026v52}. Version 6.1 is a different model: it replaces the bolometric shoreline with a cumulative-XUV retention criterion, assigns a composition class in every Monte Carlo draw, and, most importantly for the use of the output, reports lists separately for each facility and each composition class. We never merge JWST-scored and ELT-scored planets into one list, because the two facilities answer different questions and their detection probabilities are not on a common scale. Nor should the scores be read as probabilities of life, and v5.2 and v6.1 numbers should not be compared row by row.

We state the intended uses at the outset, because they determine how the outputs should be read:
\begin{itemize}
\item[\textbf{UC1}] \textbf{JWST shortlist of temperate rocky planets.} Proposers and review panels assembling rocky-planet emission or transmission programmes, of the kind recommended by \citet{Redfield2024}, can use the JWST rocky list to weigh signal strength against the chance that a target is temperate and has kept an atmosphere (Section~\ref{sec:uc1}).
\item[\textbf{UC2}] \textbf{ELT preparatory list of non-transiting planets.} Teams preparing reflected-light imaging with ELT/PCS can use the ELT rocky list to decide which known radial-velocity planets merit precursor work, such as orbit refinement and stellar characterisation (Section~\ref{sec:uc2}).
\item[\textbf{UC3}] \textbf{Rocky versus Hycean triage.} For temperate planets of intermediate size, the per-draw class probabilities show whether a target belongs in a rocky-atmosphere programme or a Hycean programme, and how strongly that verdict depends on the adopted Hycean climate criterion. This branch is secondary and indicative in v6.1 (Section~\ref{sec:uc3}).
\item[\textbf{UC4}] \textbf{Which assumption matters most.} A global Saltelli--Sobol analysis identifies the model inputs that move the lists the most, and therefore the measurements or theoretical constraints that would most change target priorities. In this paper UC4 is addressed at the global level only; a per-target value-of-information analysis is left for future work (Section~\ref{sec:sobol}).
\end{itemize}

Section~\ref{sec:method} describes the model, Section~\ref{sec:data} the input catalogue, Section~\ref{sec:validation} the checks we ran, Section~\ref{sec:results} the lists for UC1--UC3, Section~\ref{sec:sobol} the sensitivity analysis and UC4, and Section~\ref{sec:limits} the limitations. Section~\ref{sec:howto} explains how to rerun the lists, and Section~\ref{sec:concl} summarises.

\section{Framework}
\label{sec:method}

\subsection{Score, vetoes and ranking statistics}
\label{sec:score}

For every planet we generate $n = 10{,}000$ Monte Carlo realisations (draws). In each realisation the score is a weighted geometric mean of seven sub-scores in $[0,1]$: physical (size and mass), climate, atmospheric retention, stellar activity, geological regime, magnetic protection and observability,
\begin{equation}
U = \prod_{i=1}^{7} S_i^{\,w_i}, \qquad (w_1,\dots,w_7) \sim {\rm Dirichlet}(2,2,2,2,1,1,2),
\end{equation}
with the weights redrawn in each realisation so that the relative importance of the sub-scores is itself uncertain. The Dirichlet distribution is the natural prior on the probability simplex ($w_i \ge 0$, $\sum_i w_i = 1$). Its concentration vector assigns $\alpha_i = 2$ (expected weight $1/6$) to the five better-constrained terms (physical, climate, atmospheric retention, stellar activity and observability) and $\alpha_i = 1$ (expected weight $1/12$) to geological regime and magnetic protection, which lack direct observational constraints for exoplanets and rely on broader priors; the modest total concentration ($\sum_i \alpha_i = 12$) keeps the weight prior wide rather than tightly peaked. This is a pragmatic prior choice rather than one derived from first principles, and testing sensitivity to the Dirichlet concentration parameters is deferred to future work. A sub-score at or below $10^{-12}$ acts as a veto and sets $U=0$ for that realisation. Unlike the most-restrictive-criterion rule of SEPHI~2.0 \citep{RodriguezMozos2025}, this keeps a compensatory mean: fatal failures are removed by the vetoes and so lower $\Palive$, whereas a weak but non-zero sub-score lowers $U$ by the factor $S_i^{w_i}$ without deciding it alone, so a partial deficit can be offset by strengths in other sub-scores. Because the weights are redrawn in every realisation, the list statistics average over the relative importance of the sub-scores instead of resting on one fixed set of weights. Because of these vetoes the distribution of $U$ for a planet is a mixture of a point mass at zero and a continuous part, and a single summary statistic would hide that structure. We therefore report three quantities per planet and per list: the survival probability $\Palive$, the fraction of realisations with $U>0$; the conditional score, the median of $U$ over surviving realisations; and the list score
\begin{equation}
{\rm UHI}_{\rm rank} = \Palive \times {\rm median}(U \mid U>0),
\end{equation}
which is the order used for the nominal lists. In addition, for each list we record in every realisation which planets fall in the top $N$ (Section~\ref{sec:topn}).

\subsection{Climate and the transition to synchronous rotation}
\label{sec:climate}

For planets that are not tidally locked, the climate term relies on a one-dimensional energy-balance calculation \citep{North1981} with an albedo that depends on the stellar effective temperature \citep{Shields2013}, and a runaway-greenhouse veto at 282~W~m$^{-2}$ of absorbed flux, the thermal radiation limit computed by \citet{Goldblatt2013}. Version 5.2 switched to a separate parameterisation for synchronously rotating planets at a hard stellar temperature cut. In v6.1 that cut is replaced by a smooth probability of synchronous rotation,
\begin{equation}
p_{\rm sync} = \sigma\!\left(\frac{M_{\rm lock}-M_\star}{0.05\,\Msun}\right)\sigma\!\left(\frac{a_{\rm lock}-a}{0.03\,{\rm au}}\right),
\end{equation}
where $\sigma$ is the logistic function and $M_{\rm lock}\sim U(0.55,0.65)\,\Msun$ and $a_{\rm lock}\sim U(0.15,0.25)$~au are drawn per realisation. This is an assumption: it is a smoothed form of the earlier rule rather than a tidal-locking timescale. In synchronous realisations the inner edge is widened following the general-circulation results of \citet{Yang2013} and \citet{Kopparapu2016}, with an efficacy $e(P)$ that rises linearly from zero for orbital periods below $P_{\rm lo}\sim U(4,6)$~d to one above $P_{\rm hi}\sim U(15,25)$~d, motivated by the circulation regimes identified by \citet{HaqqMisra2018} and the slow-rotation transition discussed by \citet{Kopparapu2016}. The linear form is again our assumption. In these realisations the runaway veto of the one-dimensional model is not applied, and the outer edge for M-dwarf planets allows a CO$_2$ partial pressure up to a sampled maximum, following the climate modelling of \citet{Turbet2018}.

\subsection{Atmospheric retention from the cumulative XUV fluence}
\label{sec:xuv}

The retention term is where v6.1 departs most from v5.2. Rather than a bolometric shoreline \citep{Zahnle2017}, it uses the lifetime-integrated X-ray and extreme-ultraviolet fluence. For fully convective stars we follow \citet{Pass2025}: the saturated activity level is drawn as $\log R_{\rm sat}\sim\mathcal{N}(-3.05,0.06)$, the saturation time is $t_{\rm sat}=(5.9-15.4\,M_\star/\Msun)$~Gyr multiplied by a log-normal factor of width 0.3 (with a floor of 0.01~Gyr), the pre-main-sequence correction is interpolated from their Table~1, and a flare enhancement $f_{\rm flare}\sim U(1,1.5)$ brackets their adopted value of 1.25. For stars above $0.35\,\Msun$ we use the cumulative X-ray output of the Mors rotational-evolution model \citep{Johnstone2021}, run on the stellar tracks of \citet{Spada2013}, with the initial-rotation percentile drawn uniformly between 5 and 95 and the track rescaled to the observed luminosity. The two prescriptions are blended logarithmically with weight $\sigma[(M_\star-0.35\,\Msun)/0.03\,\Msun]$. The fluence $I$ is normalised by $I_E=2.0\times10^{18}$~erg~cm$^{-2}$ as in \citet{Pass2025}, and their atmosphere retention metric,
\begin{equation}
{\rm ARM} = 4\log v_{\rm esc} - \log\left(I/I_E\right) - 3.16,
\end{equation}
with $v_{\rm esc}$ in km~s$^{-1}$, measures the distance of a planet from the XUV shoreline. The retention sub-score is
\begin{equation}
S_{\rm atm} = \Phi\!\left[\frac{({\rm ARM}+c)\ln 10}{s}\right],
\end{equation}
where $\Phi$ is the standard normal cumulative distribution function, $c$ is an offset of the shoreline in dex and $s$ its width in natural-log units. The nominal values $c=0$ and $s=1$ are assumptions; both are varied in the sensitivity analysis, where they turn out to be the two most influential inputs for the rocky lists (Section~\ref{sec:sobol}). The XUV environment enters the score only once: the stellar-activity sub-score keeps its present-day stress and flare terms, and the geological term no longer carries a separate erosion flag. The present-day stress term scales the instellation by a fractional XUV luminosity that is constant until a saturation age $t_{\rm sat}$ and then declines as $(t/t_{\rm sat})^{-1.23}$, the slope of the solar-analogue fit of \citet{Ribas2005}, normalised to the Sun at 4.6~Gyr. We take $t_{\rm sat}=0.1$~Gyr for $M_\star\geq0.6\,\Msun$, as in \citet{Lopez2012}; $t_{\rm sat}=1$~Gyr for less massive stars, the use of the relation for all spectral types and the 50\% stress level at ten times the present-day Earth value (xuv\_mid) are our assumptions.

\subsection{Composition classes and branches}
\label{sec:classes}

Each realisation assigns the planet to one of three classes: R (no H$_2$ envelope; rocky or water-rich), H (Hycean; a water-rich interior beneath an H$_2$-rich atmosphere, \citealt{Madhusudhan2021}) or G (gas-rich sub-Neptune). The route by which the class is assigned depends on the available evidence.
\begin{itemize}
\item \emph{Density route} (measured radius and $M/\sigma_M>3$). Using the mass--radius curves of \citet{Zeng2019}, class R is compatible when the radius does not exceed the 50\% water curve at the planet's mass and equilibrium temperature; class H when the radius lies above the pure-silicate curve and below a maximum Hycean radius $R_{H,\max}(M)$; and class G when the radius lies above the silicate curve. $R_{H,\max}$ is interpolated linearly in $\log M$ through the values of 2.3\,$\Rearth$ at 5\,$\Mearth$ and 2.6\,$\Rearth$ at 10\,$\Mearth$ given by \citet{Madhusudhan2021}. The class is then drawn among the compatible classes with probabilities proportional to their prior weights; if no class is compatible, the planet is assigned class R. The choice of the 50\% water line and of the silicate curve as boundaries is our assumption.
\item \emph{Radius route} (radius without a usable mass). The probability of an H$_2$ envelope is $\sigma[(R_p-R_v)/w]$, with the radius-valley location $R_v=1.8197\,(M_\star/\Msun)^{0.14}\,\Rearth$ from \citet{Venturini2024} and $w\sim U(0.08,0.20)\,\Rearth$. Unmeasured masses are drawn from the class-dependent relations of \citet{Otegi2020}, with coefficients sampled within their uncertainties.
\item \emph{Mass route} (mass or minimum mass without a radius). $P(R)=\sigma[(\ln M_v-\ln M)/0.3]$ with $M_v\sim U(3,6)\,\Mearth$, a fully assumed transition; the radius then follows \citet{Otegi2020} for the drawn class.
\end{itemize}
The nominal class prior is $\pi_R:\pi_H:\pi_G=1:1:1$, applied consistently on all three routes. For radial-velocity planets the inclination is drawn isotropically, and a physical floor $R_\star\geq0.5\,R_{\rm MS}(M_\star)$, with the main-sequence relation of \citet{Demircan1991}, removes unphysical stellar-radius draws for hosts whose catalogue radius is compatible with the main sequence.

The score is then evaluated along two branches. In the \emph{rocky} branch the physical term has no upper radius limit, because that distinction is already made by the class assignment, and the retention term is the XUV criterion above. In the \emph{Hycean} branch the physical term is a trapezoid in mass with corners at 1, 2, 10 and 15\,$\Mearth$ (an assumption), the climate term requires the instellation to stay below $f_{\rm cl}\sim U(1,2)$ times the runaway limit tabulated by \citet{Innes2023} for the sampled H$_2$ pressure ($\log P_{\rm H_2}\sim U(0,2)$~bar) and stellar type, corrected for the Bond albedo, and the retention term compares the envelope loss time from energy-limited escape \citep{Owen2017} with 1~Gyr. The present-day EUV is estimated from the X-ray luminosity with the relation of \citet{SanzForcada2011} or taken from the Mors model. The Hycean climate criterion of \citet{Innes2023} is the main case; the more permissive criterion of \citet{Madhusudhan2021} is kept as a variant (Section~\ref{sec:uc3}). A planet contributes to the rocky list only in realisations in which it is class R, and to the Hycean list only in realisations in which it is class H.

\subsection{Facility-specific observability}
\label{sec:obs}

The observability term is a detection probability for the one facility suited to the planet. Transiting planets are scored for JWST: the transmission signal is scaled from the TSM and the emission signal from the ESM \citep{Kempton2018}, each predicted signal-to-noise ratio carries a log-normal model uncertainty (width 0.5 in the natural log), transmission is reduced by a cloud factor drawn between 0.5 and 1, and a systematic noise floor of 20~ppm per spectral channel, the value adopted by \citet{Greene2016} for NIRISS, is treated as independent over ten channels, so that $20/\sqrt{10}$~ppm is added in quadrature to the photon noise. The probability of detection is $1-[1-\Phi({\rm SNR}_{\rm tr}-k)][1-\Phi({\rm SNR}_{\rm em}-k)]$ for a threshold $k$. Non-transiting planets are scored for ELT/PCS with a photon-noise reflected-light model at 0.75~$\mu$m: the contrast is $A_g(R_p/a)^2/\pi$ at quadrature for a Lambertian sphere with geometric albedo $A_g=\tfrac{2}{3}A_B$, and a planet whose maximum angular separation falls inside an inner working angle drawn between 2 and 4~$\lambda/D$ is undetectable in that realisation; the probability is $\Phi({\rm SNR}-k)$. We take $k=5$ as the main threshold and $k=3$ as a variant representing deeper programmes. These observability models are deliberately simple and are no substitute for an exposure-time calculation.

\subsection{Random numbers and reproducibility}
\label{sec:rng}

The main random sequence (seed 42) is consumed in the same order as in v5.2, and the new v6.1 draws come from a separate generator seeded by a checksum of the planet name, so that every planet sees the same random numbers in every variant. Reruns in the same software environment are bit-identical.

\section{Data}
\label{sec:data}

The input is a snapshot of the Planetary Systems Composite Parameters table (\texttt{pscomppars}) of the NASA Exoplanet Archive \citep{Christiansen2025}, downloaded on 2026 November 18. Of its 6,470 rows, 6,112 have the parameters needed for scoring. Asymmetric uncertainties are sampled as split normals. The snapshot is distributed with the release together with its MD5 checksum (\texttt{f2f70b24e67a9054940267e8081c81e8}), so the lists can be regenerated exactly or recomputed on a newer snapshot.

\section{Validation}
\label{sec:validation}

We ran six checks before producing any list.
\begin{enumerate}
\item \textbf{Backward compatibility.} With all v6 modules switched off, the code reproduces the released v5.2 results for all 6,112 planets with a maximum difference of $9.2\times10^{-15}$. The v5.2 code itself, rerun in the same software environment, shows the same difference with respect to its archived output, so the residual comes from the numerical environment rather than from the new code.
\item \textbf{Synthetic Solar System planets.} Copies of the Earth, Venus and Mars placed at 10~pc behave as expected at $5\sigma$: the Earth twin survives in 99\% (non-transiting) and 100\% (transiting) of realisations and sits at the 99.87th percentile of the score distribution, while the Venus and Mars twins are vetoed in every realisation.
\item \textbf{Independent recomputation.} For GJ~273~b, Proxima~Cen~b and TRAPPIST-1~d, the instellation, the cumulative XUV fluence and its 16th--84th percentiles, the retention metric, $p_{\rm sync}$ and the class probabilities were recomputed by a separate script. Continuous quantities agree within 1.4\% and class probabilities within 0.007 in absolute terms, consistent with Monte Carlo noise.
\item \textbf{Retention metric.} Recomputing the corrected XUV fluence of \citet{Pass2025} gives $0.108/a^2$ and $0.379/a^2$ (with $a$ in au) for 0.10 and 0.30\,$\Msun$, against their tabulated $0.11/a^2$ and $0.38/a^2$; the retention metric matches hand calculations from their equations within 0.01 for 14 planets.
\item \textbf{Sensitivity-analysis code.} The Sobol estimator and report were checked on the Ishigami function \citep{Ishigami1990}, whose indices are known analytically.
\item \textbf{Automated tests.} The released code passes its full test suite, including 21 unit and regression checks for v6.1 and the v5.2-compatibility run on the whole catalogue.
\end{enumerate}

\section{Results by use case}
\label{sec:results}

At both thresholds 255 planets survive in at least one realisation of the rocky branch (167 transiting, scored for JWST, and 88 non-transiting, scored for ELT/PCS), and 71 in the Hycean branch (26 and 45). Requiring $\Palive\geq0.01$ leaves 130 rocky and 22 Hycean planets. The numbers of survivors are the same at $5\sigma$ and $3\sigma$; the threshold changes the scores and the order within each list. Absolute list scores differ systematically between the two facilities (at $5\sigma$ the leading ELT score is more than three times the leading JWST score), which is another reason to read each list only within its facility.

\subsection{Top-N membership}
\label{sec:topn}

Because the nominal order hides how often a planet actually reaches the top of its list, we recorded in each of the 10,000 realisations which planets of a given facility list occupied the top 5 and the top 10, counting a vetoed realisation as absent. $P({\rm top}\,N)$ is the fraction of realisations in which the planet is in the top $N$, and $P({\rm top}\,N\mid{\rm alive})=P({\rm top}\,N)/\Palive$ is the same fraction over the realisations in which it survives. These per-draw scores reproduce the released list statistics to $10^{-16}$, so they are the released realisations and not a new run.

\subsection{UC1: JWST rocky list}
\label{sec:uc1}

Table~\ref{tab:jwst} gives the ten transiting planets with the highest list score at $5\sigma$. The list is led by TRAPPIST-1~e, Kepler-442~b, TRAPPIST-1~f, TRAPPIST-1~d and TOI-1452~b. Three of the ten (TRAPPIST-1~e, Kepler-442~b and TRAPPIST-1~d) enter the top 10 in at least half of all realisations, with the highest value, 0.62, for TRAPPIST-1~e. Over the realisations in which they survive, all ten are in the top 10 more than half of the time. Membership is therefore limited mainly by survival: Kepler-1649~c, for instance, survives in 72\% of realisations and then reaches the top 10 in 54\% of them. The top 5 is much less stable: no planet reaches the top 5 in half of all realisations, and the planets in positions three to ten change places frequently. At $3\sigma$, which favours planets whose signal is marginal at $5\sigma$, the leading five become Kepler-442~b, TRAPPIST-1~e, Kepler-452~b, Kepler-1544~b and Gliese~12~b, and one planet (Kepler-442~b, 0.63) enters the top 10 in more than half of all realisations.

For a JWST proposer the useful information is in the columns beside the rank. A planet with high $\Palive$ but modest conditional score is a safe but unspectacular choice; a planet with high conditional score and low $\Palive$, such as TRAPPIST-1~g, is a gamble on the shoreline. Because the retention term is built on the same XUV fluence that a shoreline survey is designed to test, the outcome of such a survey (atmosphere detected or not) can be compared with the retention probabilities as a direct test of the model.

\begin{table*}
\centering
\caption{JWST (transiting) rocky list at $5\sigma$: the ten planets with the highest list score. $\Palive$ is the survival probability; $P({\rm top}\,10)$ is computed over all realisations and $P({\rm top}\,10\mid{\rm alive})$ over surviving realisations, within the JWST list only.}
\label{tab:jwst}
\begin{tabular}{rlcccc}
\toprule
Rank & Planet & UHI$_{\rm rank}$ & $\Palive$ & $P({\rm top}\,10)$ & $P({\rm top}\,10\mid{\rm alive})$ \\
\midrule
1 & TRAPPIST-1 e & 0.082 & 0.817 & 0.621 & 0.760 \\
2 & Kepler-442 b & 0.067 & 0.782 & 0.514 & 0.658 \\
3 & TRAPPIST-1 f & 0.061 & 0.382 & 0.333 & 0.870 \\
4 & TRAPPIST-1 d & 0.054 & 0.966 & 0.559 & 0.579 \\
5 & TOI-1452 b & 0.045 & 0.566 & 0.383 & 0.677 \\
6 & Gliese 12 b & 0.043 & 0.824 & 0.464 & 0.562 \\
7 & Kepler-452 b & 0.040 & 0.480 & 0.308 & 0.641 \\
8 & Kepler-1544 b & 0.040 & 0.452 & 0.294 & 0.649 \\
9 & Kepler-1649 c & 0.037 & 0.724 & 0.387 & 0.535 \\
10 & TRAPPIST-1 g & 0.036 & 0.192 & 0.171 & 0.889 \\
\bottomrule
\end{tabular}
\end{table*}

\subsection{UC2: ELT rocky list}
\label{sec:uc2}

For non-transiting planets neither the TSM nor the ESM is defined, and similarity indices do not account for angular separation or contrast. Table~\ref{tab:elt} lists the leading ELT/PCS targets at $5\sigma$: GJ~273~b, Wolf~1061~c, GJ~1061~d, Proxima~Cen~b and GJ~3378~b head the list, and the first five are the same at $3\sigma$. Four planets (GJ~273~b, GJ~1061~d, Proxima~Cen~b and Ross~128~b) enter the top 10 in at least half of all realisations, the highest being GJ~273~b at 0.55. Conditional on survival the ELT list is notably stable: nine of the ten stay in the top 10 in at least 80\% of the realisations in which they survive. What limits them is survival itself, with $\Palive$ between 0.48 and 0.61 for the five leaders, which reflects how uncertain the retention and class of these minimum-mass planets remain. None of these planets has a measured radius, so all of them are classified through the mass route, and a radius constraint or a true mass would change their standing more directly than a refinement of the ELT model.

\begin{table*}
\centering
\caption{ELT/PCS (non-transiting) rocky list at $5\sigma$, as in Table~\ref{tab:jwst}.}
\label{tab:elt}
\begin{tabular}{rlcccc}
\toprule
Rank & Planet & UHI$_{\rm rank}$ & $\Palive$ & $P({\rm top}\,10)$ & $P({\rm top}\,10\mid{\rm alive})$ \\
\midrule
1 & GJ 273 b & 0.287 & 0.569 & 0.548 & 0.964 \\
2 & Wolf 1061 c & 0.249 & 0.482 & 0.480 & 0.996 \\
3 & GJ 1061 d & 0.222 & 0.605 & 0.541 & 0.894 \\
4 & Proxima Cen b & 0.202 & 0.549 & 0.543 & 0.989 \\
5 & GJ 3378 b & 0.164 & 0.544 & 0.428 & 0.786 \\
6 & Ross 128 b & 0.163 & 0.611 & 0.534 & 0.874 \\
7 & GJ 251 b & 0.146 & 0.382 & 0.339 & 0.889 \\
8 & GJ 682 b & 0.137 & 0.288 & 0.280 & 0.973 \\
9 & GJ 433 d & 0.105 & 0.191 & 0.188 & 0.984 \\
10 & Gl 725 B c & 0.099 & 0.190 & 0.189 & 0.993 \\
\bottomrule
\end{tabular}
\end{table*}

\subsection{UC3: rocky versus Hycean triage (indicative)}
\label{sec:uc3}

The class probabilities give, for each intermediate-size planet, the fraction of realisations in which it is rocky, Hycean or gas-rich, and they show which evidence route produced that verdict. For the Hycean branch itself the results are fragile. Under the \citet{Innes2023} criterion 71 planets survive at least once, but none survives in more than about 15\% of realisations (the highest is Kepler-1630~b in the JWST list, $\Palive=0.152$), and no realisation has more than five Hycean survivors in either facility list, so every survivor is automatically in the top 5 and the top 10, $P({\rm top}\,N)$ equals $\Palive$ for both $N=5$ and $N=10$, and the ranking carries no information beyond survival. Under the more permissive criterion of \citet{Madhusudhan2021}, 1,021 planets survive, and the two Hycean top-10 lists share no planet at $5\sigma$ and one at $3\sigma$. The leading Hycean candidates at $5\sigma$ are Kepler-1630~b and Kepler-441~b for JWST, and Kapteyn~c and Wolf~1061~d for ELT/PCS. We report these lists for completeness and flag them as indicative: their membership is decided by the climate prescription rather than by the data, and Section~\ref{sec:sobol} shows that their sensitivity analysis does not converge.

\subsection{Robustness to the class prior and mass--radius choices}
\label{sec:prior}

We reran the main configuration with one change at a time: class priors favouring rocky (2:1:1) or volatile (0.5:1:1) planets, the transition mass of \citet{Chen2017}, two empirical transition masses fitted to the snapshot, minimum masses taken at face value or reweighted with the mass function of \citet{Howard2010}, the \citet{Chen2017} mass--radius relation in place of \citet{Otegi2020}, and two alternative rocky boundaries on the density route. At $5\sigma$, between 8 and 10 of the ten highest-ranked rocky planets of the main run remain in the top 10 in eight of the ten variants; with the volatile-favouring prior and the \citet{Chen2017} transition mass only 4 and 5 remain, respectively. Rank correlations over surviving planets stay at or above 0.93 in all variants. The rocky lists are therefore robust to most mass--radius choices but not to the class prior for planets without a measured radius, which is a genuine limit of what the current catalogue can support.

\section{Global sensitivity analysis (UC4)}
\label{sec:sobol}

\subsection{Design}

We sampled $D=32$ model assumptions (Table~\ref{tab:sobolpar} in Appendix~\ref{app:sobol}) with the Saltelli scheme \citep{Saltelli2002,Saltelli2010} as implemented in SALib \citep{Herman2017}, with base sample size $N=512$ and without second-order terms, which gives $N(D+2)=17{,}408$ model evaluations per detection threshold. Where a parameter sets the centre of a quantity that is still drawn in each realisation, every evaluation draws that quantity within a window of fixed half-width around the sampled centre (Table~\ref{tab:sobolpar}). These windows differ from the main-run priors of Section~\ref{sec:method}: they are narrower for the radius-valley width ($\pm0.03\,\Rearth$, against $U(0.08,0.20)\,\Rearth$), $\log P_{\rm H_2}$ ($\pm0.3$, against $U(0,2)$), the initial-rotation percentile ($\pm20$, clipped to 5--95, against $U(5,95)$) and $f_{\rm cl}$ ($\pm0.25$, against $U(1,2)$), and $f_{\rm flare}$ is held fixed within each evaluation instead of being drawn from $U(1,1.5)$. Each evaluation runs the production code with 100 realisations for each of the 1,214 planets of a fixed sensitivity sample, using common random numbers. For each branch we compute 17 outputs: the overlap of the top 10 with the baseline run, the Spearman correlation with the baseline, the number of surviving planets, and the list score of 14 reference planets. Outputs that are constant, nearly constant (fewer than 10\% of evaluations differ from the modal value) or numerically unstable ($\ST>1.5$) are excluded. Total-order indices \citep{Sobol2001} are estimated with the Jansen estimator \citep{Jansen1999} and averaged over the remaining outputs, and a 95\% interval for the mean comes from 500 bootstrap resamples. Our closure criterion is stated in terms of the half-width: a branch is considered converged when the 95\% interval of each of its ten leading parameters has a half-width below 0.05. Because a stricter reading, with the full width of the interval below 0.05, is also defensible, we report the number of leading parameters that satisfy it as well.

\subsection{Results}

\begin{figure*}
\centering
\includegraphics[width=\textwidth]{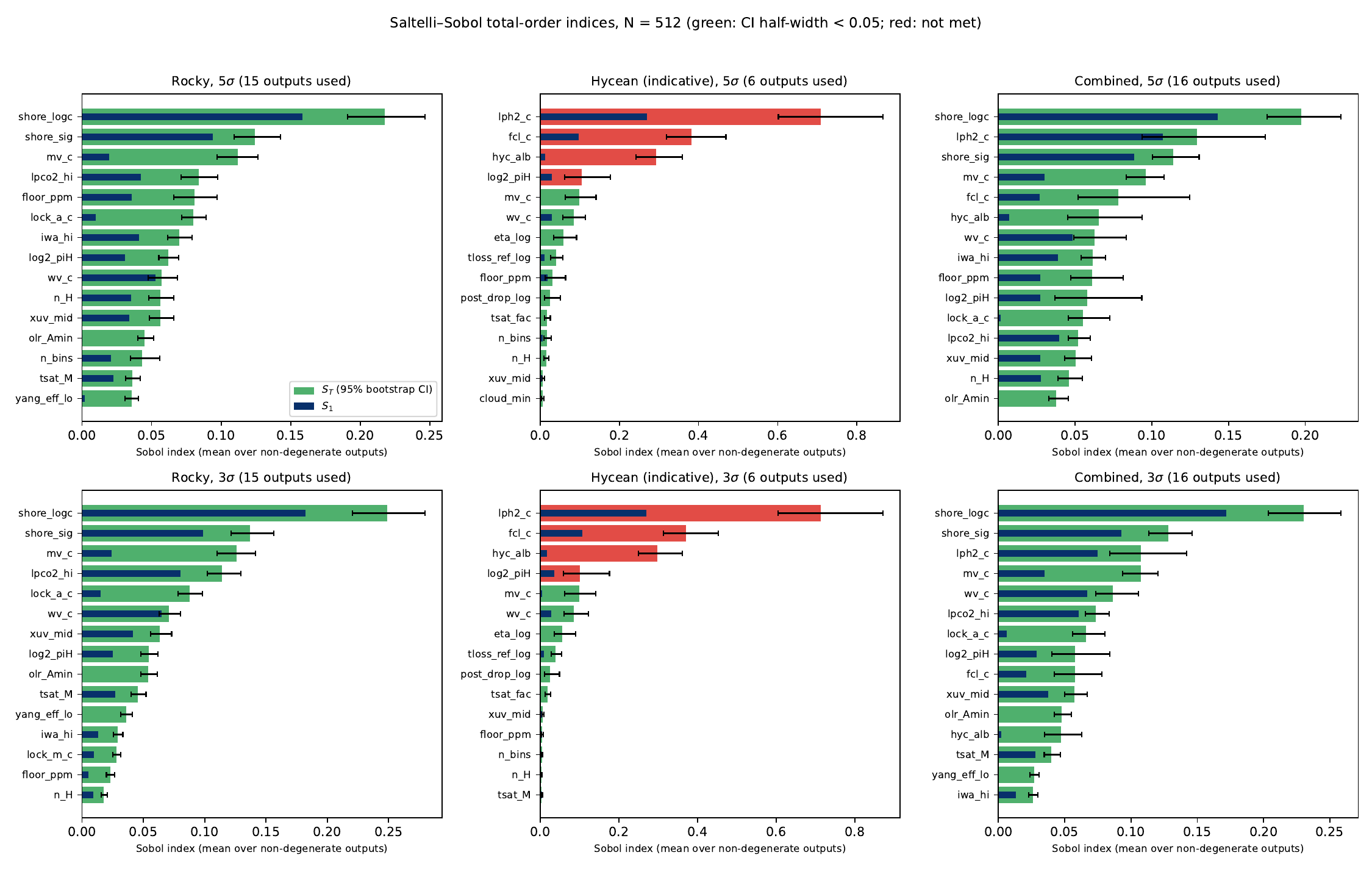}
\caption{Total-order Sobol indices $\ST$ (bars, with 95\% bootstrap intervals) and first-order indices $S_1$ (dark inner bars) for the fifteen most influential assumptions, averaged over the non-degenerate outputs of each branch, at $5\sigma$ (top) and $3\sigma$ (bottom). Green bars meet the adopted closure criterion (interval half-width below 0.05); red bars do not. The stricter full-width reading (interval width below 0.05) is met by 9 of the 10 leading rocky parameters at each threshold, 7 and 8 of 10 for the combined branch, and 2 and 3 of 10 for the Hycean branch, at $5\sigma$ and $3\sigma$ respectively; the full widths for the rocky branch are listed in Table~\ref{tab:sobol}. Parameter names are defined in Table~\ref{tab:sobolpar}.}
\label{fig:sobol}
\end{figure*}

Figure~\ref{fig:sobol} and Table~\ref{tab:sobol} summarise the results. For the rocky branch 15 of 17 outputs are usable, and all ten leading indices meet the half-width criterion at both thresholds; under the full-width reading 9 of 10 pass at each threshold, the exception being the shoreline location, whose interval is 0.055 wide at $5\sigma$ and 0.059 at $3\sigma$. The location of the XUV shoreline dominates, with $\ST=0.22$ at $5\sigma$ and 0.25 at $3\sigma$, followed by its width (0.12 and 0.14) and by the transition mass of the mass route (0.11 and 0.13). The combined branch also converges under the half-width criterion (16 usable outputs; 10 of 10 at both thresholds), while the full-width reading is met by 7 of 10 at $5\sigma$ and 8 of 10 at $3\sigma$; it is led again by the shoreline location ($\ST=0.20$ at $5\sigma$). The Hycean branch does not meet the criterion: only 6 of its 10 leading indices satisfy the half-width requirement at either threshold, and 2 of 10 ($5\sigma$) and 3 of 10 ($3\sigma$) satisfy the full-width one. Eleven of its 17 outputs are degenerate because the Hycean score vanishes in nearly every evaluation, and the estimate therefore rests on the six remaining outputs. Its ranking is led by the H$_2$ envelope pressure ($\ST=0.71$, 95\% interval 0.60--0.87), the tolerance on the Innes runaway limit (0.38) and the Hycean Bond albedo (0.29). A larger base sample cannot recover the eleven degenerate outputs, whose degeneracy comes from the model rather than from sampling noise, although it might narrow the intervals of the six retained outputs. Until the branch passes the criterion we report its ranking as indicative only.

\begin{table*}
\centering
\caption{Ten leading total-order indices for the rocky branch, with 95\% bootstrap intervals and the full width of each interval.}
\label{tab:sobol}
\begin{tabular}{lcc@{\hspace{2em}}lcc}
\toprule
\multicolumn{3}{c}{$5\sigma$} & \multicolumn{3}{c}{$3\sigma$} \\
Parameter & $\ST$ [95\% CI] & Width & Parameter & $\ST$ [95\% CI] & Width \\
\midrule
shore\_logc & 0.218 [0.191, 0.246] & 0.055 & shore\_logc & 0.249 [0.220, 0.280] & 0.059 \\
shore\_sig & 0.124 [0.110, 0.143] & 0.033 & shore\_sig & 0.137 [0.122, 0.156] & 0.035 \\
mv\_c & 0.112 [0.097, 0.126] & 0.029 & mv\_c & 0.126 [0.110, 0.141] & 0.031 \\
lpco2\_hi & 0.084 [0.071, 0.098] & 0.026 & lpco2\_hi & 0.114 [0.102, 0.129] & 0.027 \\
floor\_ppm & 0.081 [0.066, 0.097] & 0.031 & lock\_a\_c & 0.088 [0.078, 0.098] & 0.020 \\
lock\_a\_c & 0.080 [0.072, 0.089] & 0.018 & wv\_c & 0.071 [0.063, 0.080] & 0.018 \\
iwa\_hi & 0.070 [0.062, 0.079] & 0.018 & xuv\_mid & 0.064 [0.056, 0.073] & 0.017 \\
log2\_piH & 0.062 [0.055, 0.070] & 0.014 & log2\_piH & 0.054 [0.048, 0.062] & 0.014 \\
wv\_c & 0.057 [0.048, 0.069] & 0.021 & olr\_Amin & 0.054 [0.048, 0.061] & 0.014 \\
n\_H & 0.057 [0.048, 0.066] & 0.018 & tsat\_M & 0.046 [0.040, 0.052] & 0.012 \\
\bottomrule
\end{tabular}
\end{table*}

\subsection{What the sensitivity map implies for follow-up}

The answer to UC4 at the global level is clear: the inputs that would most change the rocky priority lists are not instrumental but physical, namely where the XUV shoreline lies and how sharp it is. Observations that constrain atmospheric retention on rocky M-dwarf planets, which is the aim of the shoreline survey recommended by \citet{Redfield2024} and modelled by \citet{Ih2026}, would therefore feed straight back into the lists. The third input, the transition mass used when only a minimum mass is known, points to a second lever: radii or true masses for the nearby non-transiting planets that dominate the ELT list. Instrumental assumptions (the JWST noise floor, the number of scale heights probed in transmission, the ELT inner working angle) matter, but each at a lower level. The analysis is global: it shows which assumptions move the lists as a whole, not how much a specific measurement of a specific planet would change its rank. That per-target question requires a separate value-of-information calculation on the existing realisations and is deferred to a follow-up paper.

\section{Limitations}
\label{sec:limits}

The main limitations follow from the sensitivity analysis and the validation.
\begin{itemize}
\item The location and width of the XUV shoreline dominate the rocky lists and are themselves uncertain; the lists should be updated as shoreline observations accumulate.
\item The class prior matters for planets without a measured radius (Section~\ref{sec:prior}).
\item The Hycean branch depends on the adopted climate criterion and does not converge in the sensitivity analysis; it is indicative.
\item The synchronous-rotation probability is a smoothed rule rather than a tidal-locking timescale. A comparison with the locking time of \citet{Gladman1996}, in the form used by \citet{Barnes2017}, suggests that the rule is conservative, and replacing it is planned.
\item The observability models are simplified and are no substitute for instrument simulators or exposure-time calculators.
\item Catalogue parameters are heterogeneous, and some planets survive only because of wide stellar-radius tails; reporting $\Palive\geq0.01$ alongside $\Palive>0$ limits the impact of such cases.
\item Only JWST and ELT/PCS are modelled. Extending the framework to future missions such as the Habitable Worlds Observatory and LIFE is left for future work.
\end{itemize}

\section{How to use and update the lists}
\label{sec:howto}

The release contains, for every planet and every list, $\Palive$, the conditional score and its quartiles, $P({\rm top}\,10)$, the class probabilities, the evidence route and the median value of each sub-score, which lets a user trace the reasons behind any position in a list. A JWST user selects the transiting planets from the rocky-branch results, and an ELT user the non-transiting ones. The detection threshold, the Hycean climate criterion, the class prior and the mass--radius choices are command-line options, and a full run at one detection threshold, which produces all lists, takes under three minutes on a single core. A pinned environment and a test suite are included, and the Sobol report can be rebuilt from the shipped outputs. Following the recommendations of \citet{Chen2022}, every result in this paper is tied to a versioned, citable release.

\section{Conclusions}
\label{sec:concl}

We have presented version 6.1 of an open Monte Carlo framework that turns catalogue uncertainties into facility-specific priority lists for the atmospheric characterisation of temperate exoplanets. The lists are reported separately for JWST and ELT/PCS and for rocky and Hycean planets, and each entry carries a survival probability, a conditional score and a top-$N$ probability. For the rocky branch, which is the main product, the global sensitivity analysis shows that the location and width of the XUV shoreline control the lists. The ten leading rocky total-order indices meet our closure criterion (95\% half-width below 0.05) at both thresholds, and 9 of 10 also meet the stricter full-width reading. Top-$N$ membership is limited mainly by whether a planet survives the habitability vetoes. The Hycean branch is reported as indicative. We hope these lists, and the open code that produces them, help observers decide where scarce telescope time is best spent and which measurements would most change that decision.

\section*{Data and code availability}

The code (MIT licence), the input snapshot, the per-planet results for every variant, the Sobol outputs and the figures (CC~BY~4.0) are archived on Zenodo as version 6.1.1 \citep[doi:10.5281/zenodo.24431744;][]{Viridis2026v611} and developed at \url{https://github.com/effywhitedva-ops/uhi-habitability-index}. Version 6.1.1 documents the code in English and adds the top-$N$ membership script, but it is numerically identical to version 6.1 \citep[doi:10.5281/zenodo.24430693;][]{Viridis2026v61}. The earlier v5.2 release remains available as one of the versions under the concept DOI 10.5281/zenodo.24358421 \citep{Viridis2026v52}; that DOI always resolves to the newest version, so v5.2 must be selected from the version list of the record.

\section*{Acknowledgments}
This research has made use of the NASA Exoplanet Archive, which is operated by the California Institute of Technology, under contract with the National Aeronautics and Space Administration under the Exoplanet Exploration Program.

\textit{Use of generative AI.} Two custom software agents based on large language models were used in this work. Shoggothsito implemented the author's modelling decisions in Python, ran the Monte Carlo and Saltelli--Sobol computations, wrote the test suite, generated the tables and figures from scripts and kept the run logs. Aetherios audited the methods and code against the cited literature and proposed corrections. Both assisted with literature search, and Shoggothsito drafted and language-edited the manuscript text under the author's direction. No modelling decision was made by an agent: each change was proposed, reviewed and approved by the author before it entered the code, and the approvals are recorded with dates in the release changelog. Because the agents contributed to calculation, analysis and visualisation, their output was validated as described in Section~\ref{sec:validation}: reproduction of the archived v5.2 results on all 6,112 planets, synthetic Solar System planets, an independent recomputation of key quantities for three planets, a check of the retention metric against \citet{Pass2025}, a test of the sensitivity code on the Ishigami function, and a passing automated test suite. Every reference was checked against its DOI or arXiv record. The author reviewed every step and is responsible for all of the content.

\noindent\textit{Software:} Python, NumPy \citep{Harris2020}, SciPy \citep{Virtanen2020}, pandas, SALib \citep{Herman2017}, Matplotlib, Mors \citep{Johnstone2021}.

\bibliographystyle{aasjournal}
\bibliography{refs_ojap}

\appendix
\section{Parameters of the sensitivity analysis}
\label{app:sobol}

\begin{table}[!ht]
\centering
\caption{The 32 assumptions sampled in the Saltelli--Sobol analysis, with nominal values and sampled ranges. Where a parameter is the centre of a range, the range half-width given in parentheses is kept fixed; the flare enhancement (fflare\_c) is held fixed within each evaluation.}
\label{tab:sobolpar}
\footnotesize
\begin{tabular}{llll}
\toprule
Name & Description & Nominal & Range \\
\midrule
floor\_ppm & JWST systematic noise floor (ppm) & 20 & 10--50 \\
n\_H & Number of scale heights in the transmission signal & 2 & 1--5 \\
iwa\_hi & Upper inner working angle, ELT ($\lambda/D$) & 4 & 3--6 \\
iwa\_lo & Lower inner working angle, ELT ($\lambda/D$) & 2 & 1.5--3 \\
n\_bins & Spectral channels over which the floor is averaged & 10 & 1--30 \\
xuv\_mid & Present-day XUV flux at 50\% stellar stress (present-day Earth $=1$) & 10 & 3--30 \\
tsat\_M & Saturation time in the stellar-activity term, $M_\star<0.6\,\Msun$ (Gyr) & 1 & 0.5--3 \\
lpco2\_hi & Maximum $\log_{10}P_{\rm CO_2}$ (bar), M-dwarf climate & 2 & 0--2 \\
olr\_Amin & Minimum OLR coefficient $A$ (W m$^{-2}$) & 150 & 120--190 \\
flare\_late & Flare penalty, $M_\star<0.35\,\Msun$ & 0.5 & 0.2--0.8 \\
cloud\_min & Minimum cloud factor in transmission & 0.5 & 0.1--1 \\
sig\_model & Model uncertainty of the predicted SNR (ln) & 0.5 & 0.2--1 \\
yang\_eff\_lo & Minimum 3D climate efficacy (fast rotation) & 0.5 & 0--1 \\
lock\_m\_c & Centre of $M_{\rm lock}$ ($\Msun$; $\pm0.05$) & 0.60 & 0.50--0.70 \\
lock\_a\_c & Centre of $a_{\rm lock}$ (au; $\pm0.05$) & 0.20 & 0.12--0.30 \\
prot\_lo\_c & Centre of $P_{\rm lo}$ of the 3D efficacy (d; $\pm1$) & 5 & 3--7 \\
prot\_hi\_c & Centre of $P_{\rm hi}$ of the 3D efficacy (d; $\pm5$) & 20 & 12--30 \\
log\_rsat & $\log_{10}R_{\rm sat}$ & $-3.05$ & $-3.20$ to $-2.90$ \\
tsat\_fac & Factor on the \citet{Pass2025} saturation time & 1 & 0.5--2 \\
fflare\_c & Flare enhancement of the XUV fluence & 1.25 & 1.0--1.5 \\
post\_drop\_log & $\log_{10}$ post-saturation drop (fully convective) & $-2$ & $-3$ to $-1.5$ \\
pct\_c & Centre of the initial-rotation percentile ($\pm20$) & 50 & 25--75 \\
shore\_logc & Shoreline offset $c$ (dex) & 0 & $-1$ to $1$ \\
shore\_sig & Shoreline width $s$ (ln) & 1 & 0.5--2 \\
wv\_c & Width of the radius valley ($\Rearth$; $\pm0.03$) & 0.14 & 0.05--0.25 \\
mv\_c & Transition mass $M_v$ of the mass route ($\Mearth$; $\pm1.5$) & 4.5 & 2--8 \\
log2\_piH & $\log_2$ relative prior of the Hycean class & 0 & $-1$ to $1$ \\
lph2\_c & Centre of $\log_{10}P_{\rm H_2}$ (bar; $\pm0.3$) & 1 & 0.3--1.7 \\
fcl\_c & Tolerance on the \citet{Innes2023} limit ($\pm0.25$) & 1.5 & 0.8--2.5 \\
eta\_log & $\log_{10}$ factor on the escape efficiency & 0 & $-1$ to $1$ \\
tloss\_ref\_log & $\log_{10}$ reference envelope loss time (Gyr) & 0 & $-1$ to $0.65$ \\
hyc\_alb & Hycean Bond albedo & 0.12 & 0.05--0.60 \\
\bottomrule
\end{tabular}
\end{table}

\end{document}